\documentclass[aps,prl,twocolumn, titlepage,showpacs]{revtex4}

\usepackage{graphicx}
\usepackage{color}
\usepackage{dcolumn}

\usepackage{amsfonts}

\usepackage{bm}

\begin{document}

\title{Continuous and discontinuous transitions in the depinning of two-dimensional dusty plasmas on a one-dimensional periodic substrate}

\author{L. Gu$^1$, W. Li$^1$, C. Reichhardt$^2$, C. J. O. Reichhardt$^2$, M. S. Murillo$^3$, and Yan Feng$^1$ $^4$ $\ast$}
\affiliation{
$^1$ Center for Soft Condensed Matter Physics and Interdisciplinary Research, School of Physical Science and Technology, Soochow University, Suzhou 215006, China\\
$^2$ Theoretical Division, Los Alamos National Laboratory, Los Alamos, New Mexico 87545, USA\\
$^3$ Department of Computational Mathematics, Science and Engineering, Michigan State University, East Lansing, Michigan 48824, USA\\
$^4$ National Laboratory of Solid State Microstructures, Nanjing University, Nanjing 210093, China\\
$\ast$ E-mail: fengyan@suda.edu.cn}

\date{\today}

\begin{abstract}

Langevin dynamical simulations are performed to study the depinning dynamics of two-dimensional dusty plasmas on a one-dimensional periodic substrate. From the diagnostics of the sixfold coordinated particles $P_6$ and the collective drift velocity $V_x$, three different states appear, which are the pinning, disordered plastic flow, and moving ordered states. It is found that the depth of the substrate is able to modulate the properties of the depinning phase transition, based on the results of $P_6$ and $V_x$, as well as the observation of hysteresis of $V_x$ while increasing and decreasing the driving force monotonically. When the depth of the substrate is shallow, there are two continuous phase transitions. When the potential well depth slightly increases, the phase transition from the pinned to the disordered plastic flow states is continuous, however, the phase transition from the disordered plastic flow to the moving ordered states is discontinuous. When the substrate is even deeper, the phase transition from the pinned to the disordered plastic flow states changes to discontinuous. When the substrate further increases, as the driving force increases, the pinned state changes to the moving ordered state directly, so that the disordered plastic flow state disappears completely.

\end{abstract}

\maketitle

\section{I.~Introduction}

Many driven systems can be characterized by a collection of interacting point particles that passes through disordered or ordered substrates under a uniform force~\cite{Schwarz:2001, Reichhardt:2017}. Examples of these systems include vortex lattices in superconductors with periodic arrays of pinning sites~\cite{Reichhardt:1998}, arrays of nanostructured pinning sites~\cite{Harada:1996}, colloidal monolayers driven across ordered surfaces~\cite{Bohlein:2012}, Wigner crystals~\cite{Williams:1991}, and pattern-forming systems~\cite{Reichhardt:2003,Sengupta:2010}. It was discovered that these systems could exhibit critical depinning transitions~\cite{Schwarz:2001} when an applied uniform force is combined with a substrate. When the external driving force, $F_d$, is too small to overcome the confinement by the substrate, the system is trapped in one of many possible metastable configurations. As the external driving force, $F_d$, gradually increases, the initial configuration becomes unstable and moves, and may be stopped frequently by the elastic forces. As $F_d$ increases further, the system can ``avalanche'' and move at higher speeds~\cite{Schwarz:2001}. The transitions between these different states can be characterized by the critical depinning thresholds~\cite{Schwarz:2001, Reichhardt:2017}.

Dusty plasma, a collection of highly charged micron-sized dust particles in a partially ionized gas~\cite{Thomas:1996, L:1996, Konopka:2000, Merlino:2004, Fortov:2005, Morfill:2009, Bonitz:2010, Melzer:2013, Thomas:2016}, can also be coupled to a substrate, as studied in~\cite{Li:2018, Wang:2018, Li:2020}. Under typical laboratory conditions, these dust particles are charged to a high negative charge of $\sim -10^4 e$, and they can self-organize into a single layer plane, i.e., forming a two-dimensional dusty plasma (2DDP)~\cite{Feng:2011, Qiao:2014}. In experiments, these highly charged dust particles are strongly coupled, exhibiting collective solid-like~\cite{Feng:2008,Hartmann:2014, Melzer:1996} or liquid-like behaviors~\cite{Chan:2007,Feng:2010, Thomas:2004}. Substrates have been experimentally realized in 2DDPs using a striped electrode, as demonstrated in~\cite{LiF:2009, LiF:2010}. Recently, the coupling of 2DDP with a one-dimensional periodic substrate (1DPS) has been studied using simulations, which focused on the phonon spectra~\cite{Li:2018}, the structural transitions~\cite{Wang:2018}, and also the diffusion~\cite{Wang:2018, Li:2020}. If a uniform force is applied on all particles of 2DDP with 1DDP, for example using the laser radiation force in experiments~\cite{Melzer:2000}, the depinning dynamics can be investigated. The depinning dynamics of one row of dust particles in each potential well of 1DPS was investigated in~\cite{Li:2019} using simulations, and it is found that, for a certain range of the substrate depth, three different states appear as the external force increases gradually from zero, which are the pinned, disordered plastic flow and moving ordered states. However, for different configurations of 2DDP under 1DPS, such as for two rows of dust particles in each potential well of 1DPS, the depinning dynamics would be more complicated, as studied here.

A subsequent question is whether the depinning process of 2DDP on 1DPS exhibits continuous or discontinuous phase transitions. When the system interacts with the substrate, the different depinning dynamical phases can be identified by the structural symmetry, while the breaking of these symmetries can characterize these phase transitions~\cite{Reichhardt:2017}. The phase transitions might be second-order or continuous in nature, or they might be first-order and be accompanied by hysteresis~\cite{Reichhardt:1997, Reichhardt:2005}. It is also possible for various mixed first-order and second-order transitions or simple crossover behaviors to exist~\cite{Reichhardt:2017}. For the depinning dynamics of 2DDP under 1DPS, as the depth of the substrate increases, the phase transition mechanism might change from one form to another, as we study here.

The rest of this paper is organized as follows. In Sec.~II, we briefly introduce our Langevin dynamical simulation method to mimic 2DDP under 1DPS while subjected to a driving force $F_d$. In Sec.~III, we present the structural and dynamical measures of our system, including the collective drift velocity $V_x$ along the direction of the driving force, the measurement of particle structural stability $P_6$, the hysteresis of $V_x$, and the kinetic temperature. Finally, in Sec.~IV, we present a brief summary.

\section{II.~Simulation methods}

Without substrates, traditionally, 2DDP can be characterized by two dimensionless parameters~\cite{Ohta:2000, Sanbonmatsu:2001}, the coupling parameter $\Gamma = Q^2/(4 \pi \epsilon_0 a k_B T)$ and the screening parameter $\kappa \equiv a / \lambda_D$. Here, $a = (\pi n)^{-\frac{1}{2}}$ is the Wigner-Seitz radius~\cite{Kalman:2004} with areal number density $n$, $T$ is the particle kinetic temperature, $Q$ is the charge of each dust particle, and $\lambda_D$ is the screening length. 

We use Langevin dynamical simulations to investigate the depinning dynamics of 2DDP on a 1DPS, using the equation of motion~\cite{Li:2018, Wang:2018, Li:2019, Donko:2010}
\begin{equation}\label{LDE}
{	m \ddot{\bf r}_i = -\nabla \Sigma \phi_{ij} - \nu m\dot{\bf r}_i + \xi_i(t)+{\bf F}_s + {\bf F}_d,}
\end{equation}
for the dust particle $i$. Here, the first term on the right of Eq.~(\ref{LDE}) $-\nabla \Sigma \phi_{ij}$ is the binary Yukawa interaction~\cite{Liu:2003} between dust particles, $\phi_{ij} = Q^2 $exp$({-r_{ij}}/ { \lambda_D })/ {4 \pi {\epsilon_0}{r_{ij}}}$, where $r_{ij}$ is the distance between dust particles $i$ and $j$. The terms of $- \nu m\dot{\bf r}_i$ and $\xi_i(t)$ are the frictional drag and the Langevin random kicks~\cite{FengY:2008, van:1982}, respectively. We assume that the 1DPS has the form of
\begin{equation}\label{1DPS}
{	U(x) = U_0 \cos(2 \pi x/w),}
\end{equation}
so that the force from the 1DPS is ${\bf F}_s = - \frac{\partial U(x)}{\partial x}{\hat {\bf x}} = (2 \pi U_0 / w)\sin(2 \pi x /w){\hat {\bf x}}$, which is in the $x$ direction. Here, $U_0$ and $w$ are the depth and width of the potential well, in units of $ E_0 = Q^2/{4 \pi {\epsilon_0} a }$ and $b$ , respectively. The last term on the right of Eq.(1), $F_d$, is the external driving force, in units of $F_0 = Q^2/{4 \pi {\epsilon_0} a^2  }$. Note that, we use the inverse nominal 2D dusty plasma frequency, ${\omega}_{pd}^{-1} = (Q^2/2\pi\epsilon_0 m a^3)^{-1/2}$, to normalize the time scale, and use either the Wigner-Seitz radius $a$ or the lattice constant $b$ to normalize the length scale~\cite{Ohta:2000,Kalman:2004}.

We simulate $N=1024$ particles constrained within a $61.1a \times 52.9a$ 2D plane with periodic boundary conditions. Since the size in the $x$ direction is $61.1a = 32.07b$, to satisfy the periodic boundary conditions, we specify the width of the potential well as $w/b = 2.004$, which corresponds to 16 full potential wells. For the depth of the potential well, we consider four different values, $U_0/E_0=0$, $0.01$, $0.05$, $0.10$ and $0.25$. To reduce the temperature effect on the depinning dynamics, we fix the conditions of the simulated 2DDP at $\Gamma = 1000$ and $\kappa = 2$, corresponding to the typical solid/crystal state in the absence of substrates or external forces~\cite{Hartmann:2005}. The gas damping rate is chosen to be comparable to the typical experimental value of $ \nu / \omega_{pd}^{-1} = 0.027 $. For each simulation run, we integrate $ \ge 10^7$ steps of Eq.~(\ref{LDE}) with a time step of $0.0028 \omega_{pd}^{-1}$ (or $0.0007 \omega_{pd}^{-1}$ only for $U_0/E_0=0.25$) to obtain the positions and velocities of all particles. We also performed a few test runs with 4096 particles to verify that all results reported here are not affected by the total particle number. Other simulation details are the same as in~\cite{Li:2019}.

\section{III.~Results}

\begin{figure*}[htb]
	\centering
	\includegraphics{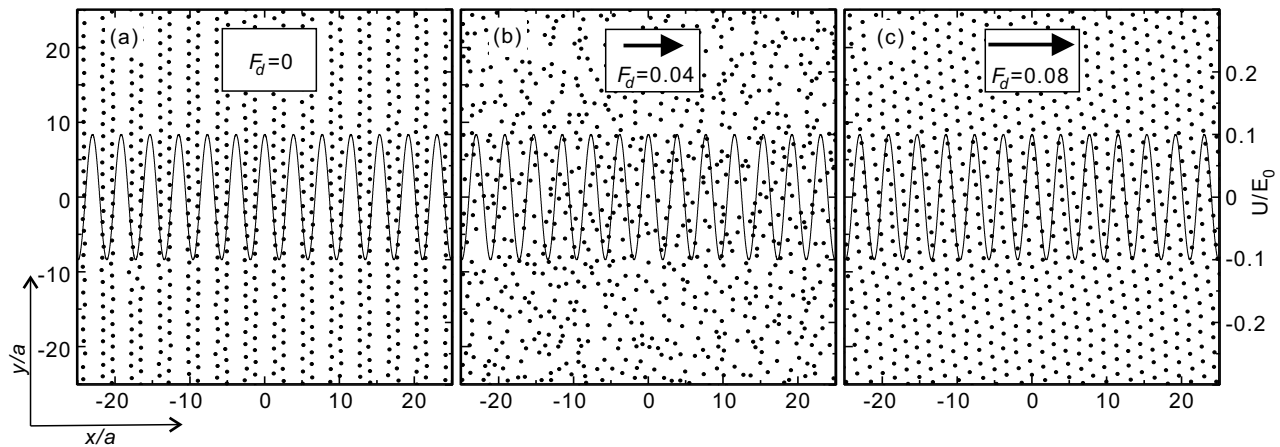}
	\caption{\label{fig:position} Snapshots of particle positions (dots) for a 2D Yukawa crystal with $\Gamma = 1000$ and $\kappa = 2$ under the 1DPS (curve) of $U(x) = U_0\cos(2\pi x/w)$ ($U_0/E_0 = 0.10 $ and period $w = 2.004b$) while experiencing different external driving forces. When $F_d/F_0 = 0$ in (a), the system is in the pinned state, so that the particles are neatly arranged in two rows within one potential well of the substrate. When $F_d/F_0 = 0.04$ in (b), the system is in the disordered plastic flow state. When $F_d/F_0 = 0.08$ in (c), the system is in the moving ordered state, so that the particles are distributed in an ordered triangular lattice, independent of the locations of the potential wells.
}
\end{figure*}

In the depinning procedure of 2DDP on 1DPS, when the driving force increases gradually, three typical dynamical states appear, which are the pinned state, the disordered plastic state and the moving ordered state, respectively, as shown in Fig.~1. When the driving force is very small, all of the particles are pinned around their equilibrium locations due to the 1DPS, so that the particles are neatly arranged in two rows within one potential well of the substrate. When the driving force is larger, some particles can escape from the 1DPS and the cages formed by their neighbors, forming a disordered plastic state. When the driving force on each particle is high enough to overcome the 1DPS, all particles move with a constant rate of increase in the velocity along the direction of the driving force, and these particles are distributed in an ordered triangular lattice, independent of the potential wells. Note that these three states are similar to the states observed for superconducting vortices~\cite{Thorel:1973}, a defective flux-line lattice~\cite{Shi:1991}, vortex lattices~\cite{Koshelev:1994}, Skyrmions~\cite{Reichhardt:2015}, and the depinning of 2DDP with only one row of particles within one potential well~\cite{Li:2019}.

\subsection{a.~Continuous and discontinuous phase transitions}

\begin{figure}[htb]
\centering
\includegraphics{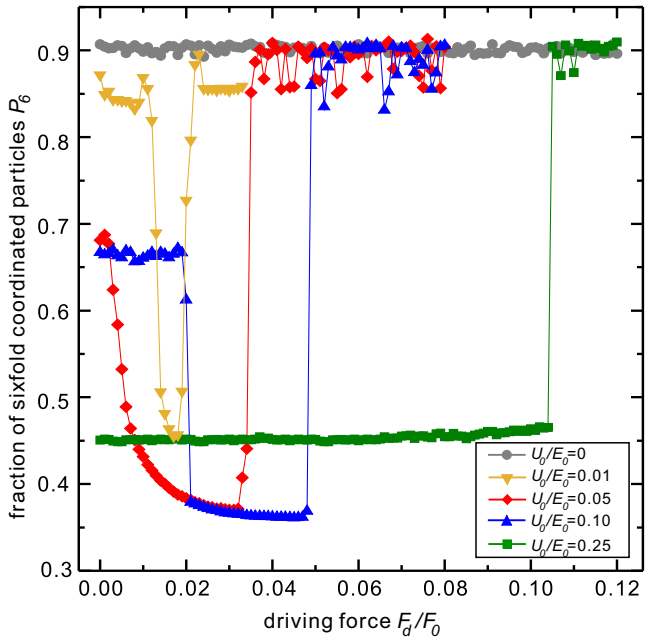}
\caption{\label{P6} (Color online). The fraction of sixfold coordinated particles~\cite{Reichhardt:2005} $P_6$ as the external driving force $F_d$ increases from zero for different substrate depths of $U_0/E_0 = 0 $, $0.01 $, $0.05 $, $0.10$, and $0.25 $. For a perfect triangular lattice, $P_6 = 1.0$, while $P_6$ is reduced for a more disordered system, since $P_6$ is defined as $\langle \sum_{i=1}^{N_d} \delta (6- z_i ) \rangle / N_d $, where $z_i$ is the coordination number of particle $i$ obtained from a Voronoi construction. When $U_0/E_0 = 0.01$, $0.05$ and $0.10$, as $F_d$ increases from 0, three different states can be clearly observed, which are the initial high value of $P_6$ (well above 0.6), then a low value of around 0.45 or even lower, and finally the high value again (above 0.8). For $U_0/E_0 = 0.05$ and $0.10$, these three values of $P_6$ correspond to the pinned, disordered plastic flow and moving ordered states observed in Fig.~1. However, when $U_0/E_0 = 0.25 $, as $F_d$ increases from 0, the value of $P_6$ jumps from the initial value of $0.45$ directly to around $0.9$. This structure measure also reflects the property of the phase transition. For $U_0/E_0 = 0.01 $, as $F_d$ increases from 0, $P_6$ drops continuously to about 0.45 and then continuously returns to its initial high value, suggesting that the two phase transitions are both continuous. Similarly, for $U_0/E_0 = 0.05$, as $F_d$ increases from 0, $P_6$ drops continuously to about 0.35 and returns suddenly to a high value of $> 0.8$, suggesting that the the first phase transition is continuous, while the second is discontinuous. For $U_0/E_0 = 0.10 $ and $0.25$, all of the phase transitions are discontinuous.
}
\end{figure}

We calculate the static structural measure of the sixfold coordinated particles $P_6$ as the driving force $F_d$ increases for five different values of $U_0/E_0$, as shown in Fig.~2. Here, $P_6$ is defined as $P_6 = \langle \sum_{i=1}^{N_d} \delta (6- z_i ) \rangle / N_d $~\cite{Reichhardt:2005}, where $z_i$ is the coordination number of particle $i$ obtained from the Voronoi construction. For a perfect triangular lattice, $P_6 = 1.0$, while, for a more disordered system, the value of $P_6$ is reduced.

We find that in our simulated system, the depinning dynamic state depends on not only the magnitude of the driving force, but also the depth of the substrate, as shown in Fig.~2. When $U_0/E_0 = 0.05$ and $0.10$, from Fig.~2, as the driving force increases from zero, the value of $P_6$ varies over three distinctive ranges, which correspond to the pinned, disordered plastic flow and moving ordered states in Fig.~1. When the driving force is very small, for the two depths of the substrate $U_0/E_0 = 0.05$ and $0.10$, $P_6 \approx 0.65$, corresponding to the pinned state. Here, within each potential well, the particles are pinned around the bottom to form two rows, as shown in Fig.~1(a). As the driving force increases gradually, the value of $P_6$ decreases substantially to a lower value of around 0.35, which is a typical value for a disordered plastic flow state. Thus, the lower value of $P_6$ of around 0.35 for $U_0/E_0 = 0.05$ and $0.10$ in the middle range of the driving force in Fig.~2 correspond to the disordered plastic flow state in Fig.~1(b). When the driving force increases enough to completely overcome the constraint from the 1DPS, $P_6$ suddenly increases to a higher value of around 0.9, corresponding to the moving ordered state in Fig.~1(c).

When the substrate depth is fairly deep, for example $U_0/E_0 = 0.25$ as shown in Fig.~2, we find that the second disordered plastic flow state disappears completely. As the driving force $F_d$ gradually increases from 0, the value of $P_6$ stays around the initial low value of about $0.45$ until $F_d/F_0 > 0.10$, then suddenly jumps directly to around 0.9. We do not find a decrease in the $P_6$ from our data analysis, suggesting that the second disordered plastic flow state is not present. Note that the initial value of $P_6$ is lower than what is found for shallower substrate depths because the 1DPS can more strongly distort the arrangement of the particles.

When the substrate depth is very shallow, for example $U_0/E_0 = 0.01$ as shown in Fig.~2, the process is slightly different. In the initial state, we find a fairly high $P_6 \approx 0.85$. This is completely different from what we observe in the pinned state for other substrates because the shallow potential well can only exert a weak constraint on the particles. When the external force increases, the stability of the system is destroyed and $P_6$ decreases to a low value of around 0.45, corresponding to the disordered plastic flow state. As the external force further increases, $P_6$ increases back to about 0.85, which suggests that the particles have rearranged into an ordered triangular lattice, independent of the potential well locations. Note that in the initial and final moving ordered states, the value of $P_6$ is nearly unchanged and is given by $P_6 \approx 0.85$.

From Fig.~2, the continuous/discontinuous property of the phase transition is visible from the variation of the value of $P_6$. For $U_0/E_0 = 0.01$, as $F_d$ increases from 0, $P_6$ diminishes continuously to about 0.45 and then continuously returns to its initial high value, suggesting that these two phase transitions are both continuous. Similarly, for $U_0/E_0 = 0.05$, as $F_d$ increases from 0, $P_6$ drops continuously to about 0.35 and then suddenly returns to a high value of $> 0.8$, suggesting that the first phase transition is continuous while the second transition is discontinuous. For $U_0/E_0 = 0.10$ and $0.25$, all of the phase transitions shown in Fig.~2 are discontinuous.

\begin{figure}[htb]
\centering
\includegraphics{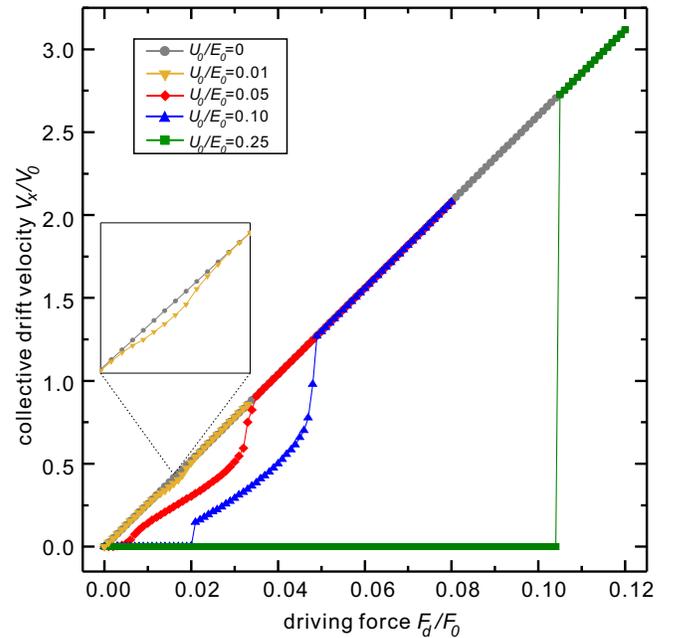}
\caption{\label{fig:Vx}(Color online). The collective drift velocity $V_x = N_d^{-1} \langle \sum_{i=1}^{N_d}{\bf v}_i \cdot \hat {\bf x}\rangle$, as the external driving force $F_d$ increases from zero for different substrates of $U_0/E_0 = 0 $, $0.01 $, $0.05 $, $0.10$, and $0.25 $. The units of $V_x$ are $V_0 = (Q^2/4\pi\epsilon_0 m a)^{1/2}$. For the shallowest substrate $U_0/E_0 = 0.01 $, we find that the collective drift velocity increases continuously and almost overlaps with the zero-substrate curve, showing only a small deviation when $F_d/F_0 \approx 0.18$. This feature suggests that the corresponding depinning involves the collective motion of all the particles, which is quite different from the plastic depinning transition. For the substrate with $U_0/E_0 = 0.05 $, when the external driving force increases gradually, $V_x$ increases continuously from the initial pinned state to the second disordered plastic flow state, and then increases discontinuously or suddenly to the final moving ordered state. For the substrate with $U_0/E_0 = 0.10$, $V_x$ increases discontinuously from the initial pinned state to the second disordered plastic flow state, and then increases discontinuously to the final moving ordered state. For the substrate with $U_0/E_0 = 0.25$, $V_x$ increases discontinuously from the initial pinned state directly to the final moving ordered state, without passing through the disordered plastic flow state. The phase transition properties exhibited by the collective drift velocity $V_x$ are consistent with those found from the static structure measure $P_6$ in Fig.~2 above.
}
\end{figure}

Our results on the collective drift velocity $V_x$ for various substrates as a function of the external driving force are presented in Fig.~3. Here, we calculate the collective drift velocity using $V_x = N_d^{-1} \langle \sum_{i=1}^{N_d}{\bf v}_i \cdot \hat {\bf x}\rangle$. The unit of $V_x$ is $V_0 = (Q^2/4\pi\epsilon_0 m a)^{1/2}$. For the typical values of $U_0/E_0 = 0.05$ and 0.10, when the external force is small, the collective drift velocity is almost zero, corresponding to the pinned state. As the external force gradually increases, the collective drift velocity increases relatively steeply, corresponding to the disordered plastic flow state. Finally, when the external force is very large, the collective drift velocity increases linearly with $F_d$, corresponding to the moving ordered state~\cite{Li:2019}. Note that, as found in~\cite{Li:2019}, for the final moving ordered state, the collective drift velocity $V_x$ also increases linearly with the external driving force $F_d$ at a fixed slope of $\nu m$, independent of the 1DPS.

Our previous conclusion about the continuous/discontinuous property of the phase transition observed from the static structural measure of $P_6$ above is further verified by the collective drift velocity $V_x$ results in Fig.~3. There are three types of dynamical states in Fig.~3. Two of them can be easily identified as the initial pinned state where the collective drift velocity is zero, and the final moving ordered state where the collective drift velocity increases linearly with $F_d$. Other data points between these two lines belong to the plastic flow state.

For the substrate with $U_0/E_0 = 0.05$, as the driving force increases gradually, the increase of $V_x$ from the initial pinned state to the second disordered plastic flow state is continuous, while the later increase of $V_x$ from the disordered plastic flow phase to the final moving ordered state is discontinuous or abrupt. For the substrate with $U_0/E_0 = 0.10$, the two-step increases of $V_x$ from the initial pinned state to the disordered plastic flow state, and then to the final moving ordered state, are both discontinuous as a function of increasing driving force. For the deep substrate of $U_0/E_0 = 0.25$, $V_x$ remains zero until the driving force increases to more than 0.1, then $V_x$ suddenly jumps directly from 0 to the final linear range with increasing $F_d$, i.e., directly from the initial pinned to the final moving ordered state, without passing through the disordered plastic flow state. Note that in addition to the static structural measures and the collective drift velocity, we provide the trajectories of our simulated 2DDP under this 1DPS, as presented in the Supplementary Materials of~\cite{SM}.

For the shallow substrate of $U_0/E_0 = 0.01$, we find that the collective drift velocity always increases continuously and almost overlaps with that of the zero-substrate case, with only a small deviation when $F_d/F_0 \approx 0.18 $, as magnified in the inset of Fig.~3. This feature suggests that for this sample the depinning process involves collective motion of all of the particles particles, which is quite different from the typical plastic depinning process from the initial pinned state for the other cases studied here. Based on the combination of the linear increase of $V_x$ with $F_d$ in the initial state in Fig.~3 with the corresponding $P_6$ result in Fig.~2 for $U_0/E_0 = 0.01$, we determine that all of the particles begin to move in the direction of the driving force as a rigid object, as the final moving ordered state. This initial state is transient, however, since its structure and collective drift velocity are partially modified when $F_d/F_0 \approx 0.18 $.

Clearly, the continuous/discontinuous property of $V_x$ at each transition observed from Fig.~3 is consistent with that presented in $P_6$ in Fig.~2. Our conclusion is based on both the static structural and dynamical measures from our simulations of 2DDP on the 1DPS.

\subsection{b.~Hysteresis of the collective drift velocity }

\begin{figure}[htb]
	\centering
	\includegraphics{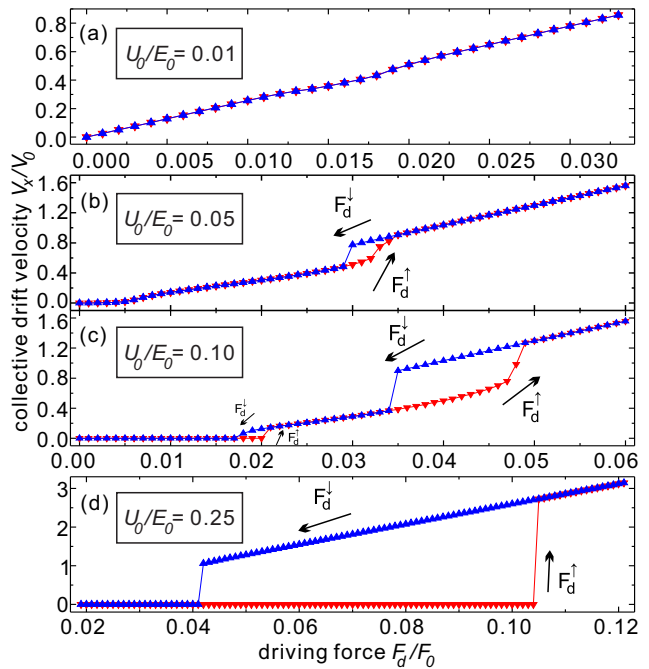}
	\caption{\label{fig:loop} (Color online). The hysteresis of the collective drift velocity $V_x$ as the external driving force $F_d$ increases and decreases monotonically, for different substrate depths of
$U_0/E_0 = 0.01 $, $0.05$, $0.10$, and $0.25$. When $F_d$ increases and decreases monotonically, our focus is whether this discontinuous phase transition will exhibit hysteresis. When $U_0/E_0 = 0.01 $ in (a), there is no hysteresis around either of the two continuous phase transitions. When $U_0/E_0 = 0.05 $ in (b), the hysteresis around $F_d/F_0 = 0.03$ is associated with the discontinuous phase transition. However, when $F_d/F_0 \approx 0.005 $, there is no hysteresis, suggesting that the phase transition from the pinned to the disordered plastic flow state is continuous. When $U_0/E_0 = 0.10 $ in (c), significant hysteresis appears near the two discontinuous phase transitions. When $U_0/E_0 = 0.25 $ in (d), we find a single large hysteresis loop, suggesting that there is only one discontinuous phase transition from the pinned to the moving ordered state.
          }
\end{figure}

For a physical procedure, the hysteresis generally results from the lagging of the system response to an external modification. Typically, a process with hysteresis shows an overshoot during its evolution~\cite{Ooyen:1992}. As described in~\cite{Schwarz:2001,Schwarz:2003, Reichhardt:2017}, when an overshoot is present, the critical depinning threshold is reduced, so that an originally non-hysteretic depinning transition becomes increasingly hysteretic. Here, we investigate whether the hysteresis feature exists in the depinning of 2DDP with 1DPS.

From~\cite{Reichhardt:1998, Reichhardt:1997, Schwarz:2001}, the hysteresis feature is directly related to the property of the depinning phase transition. For the first-order, or discontinuous, phase transition, when the static structural or dynamical measures both show abrupt jumps, hysteresis would in principle be expected to appear, whereas for a second-order, or continuous, phase transition, there should not be any hysteresis. We next present the overall drift velocity $V_x$, as the driving force $F_d$ increases and decreases monotonically in our simulations.

Our results on the hysteresis of the collective drift velocity as the driving force increases and decreases monotonically from our simulations, are presented in Fig.~4. Here, ${F_d}^{\uparrow}$ represents the increase of $F_d/F_0$ from $0$ to $0.125 $, while ${F_d}^{\downarrow}$ represents the decrease of $F_d/F_0$ from $0.125 $ back to $0$. When $F_d$ increases and decreases monotonically, our focus is on whether the phase transition would exhibit hysteresis. For $U_0/E_0 = 0.01 $ in Fig.~4(a), as $F_d$ either increases or decreases monotonically, $V_x$ always follows the same trace without any hysteresis. For $U_0/E_0 = 0.05 $ in Fig.~4(b), there is a single hysteresis loop when $F_d/F_0$ is around $0.03$, corresponding to the transition between the disordered plastic flow and the moving ordered state. For $U_0/E_0 = 0.10 $ in Fig.~4(c), there are two obvious hysteresis loops. A smaller loop is centered at $F_d/F_0 \approx 0.02$, while the larger loop is near $F_d/F_0 \approx 0.04$. For $U_0/E_0 = 0.25$ in Fig.~4(d), there is a single huge hysteresis loop when $0.4 \le F_d \le 0.10$, corresponding to the transition between the pinned state and the moving ordered state.

As our chief conclusion in this paper, we discover both first-order and second-order depinning phase transitions in the 2DDP under 1DPS from our simulations. The first-order depinning transition exhibits a discontinuity in the structural/dynamical measures when the external driving force increases, and hysteresis appears when the driving force decreases/increases monotonically. However, for the second-order depinning transition, the structural/dynamical measures are always continuous, and there is no hysteresis.

Clearly, both the hysteresis in Fig.~4 and the discontinuities in $P_6$ and $V_x$ in Figs.~2 and 3 indicate that the transition from the plastic flow to the moving ordered state for the substrate with $U_0/E_0 = 0.05$, the two transitions for the substrate with $U_0/E_0 = 0.10$, and the single transition for the substrate with $U_0/E_0 = 0.25$ are first-order. The lack of hysteresis in Fig.~4 and the continuity of $P_6$ and $V_x$ in Figs.~2 and 3 indicate that the two transitions for the substrate with $U_0/E_0 = 0.01$ and the transition from the pinned state to the plastic flow state for the substrate with $U_0/E_0 = 0.05$ might be second-order.

\subsection{c.~The property of kinetic temperature}

\begin{figure}[htb]
\centering
\includegraphics{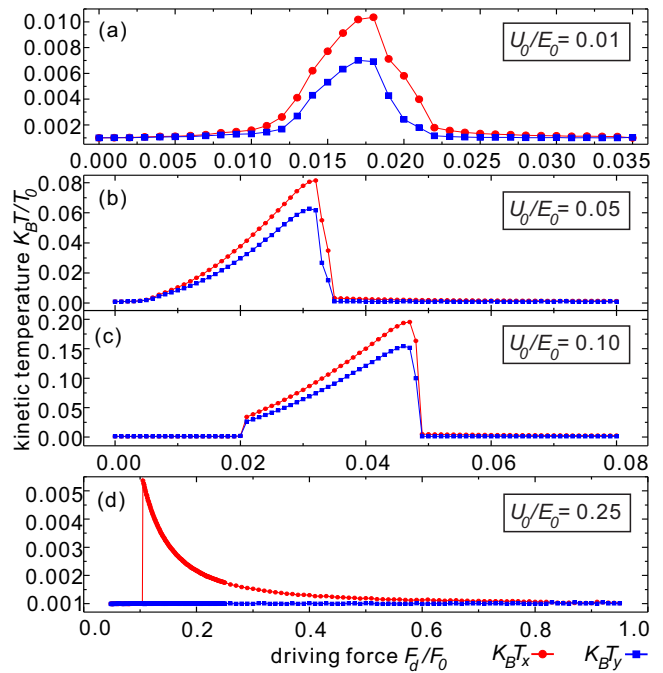}
\caption{\label{fig:kbt}(Color online). The kinetic temperature $k_BT_x$ and $k_BT_y$ as a function of increasing driving force $F_d$ for different substrate depths of $U_0/E_0 = 0.01 $, $0.05$, $0.10$, and $0.25$. The kinetic temperature is calculated using $k_BT = m \langle \sum_{i=1}^{N}\left({\bf v}_i - \overline {\bf v}\right)^{2}\rangle/2$, so that the collective drift motion is removed. The kinetic temperature is in the units of $T_0 = Q^2/4\pi\epsilon_0 a$. For the substrate with $U_0/E_0 = 0.01 $, the kinetic temperature change at the two depinning transitions is continuous. For $U_0/E_0 = 0.05 $ in (b), we find that the kinetic temperature of the first depinning transition from the pinned to the disordered plastic flow state increases continuously, while at the second depinning transition from the disordered plastic flow phase to the moving ordered state, both $k_BT_x$ and $k_BT_y$ drop abruptly. For $U_0/E_0 = 0.10 $ in (c), there are two abrupt jumps of kinetic temperature at the two depinning transitions. For $U_0/E_0 = 0.25 $ in (d), the variation of $k_BT_x$ and $k_BT_y$ are no longer synchronized, unlike what is found in (a), (b), and (c). We find that, when $F_d/F_0 \approx 0.1$ and the initial pinned state transitions directly to the moving orderly state, the value of $k_BT_x$ increases steeply to its maximum, and then decays gradually as $F_d$ increases further; however, the value of $k_BT_y$ does not change substantially.
  }
\end{figure}

To explore the underlying dynamical characteristics of the depinning phase transition of 2DDP under 1DPS, we also calculate the kinetic temperature in Fig.~5 using $k_BT = m \langle \sum_{i=1}^{N}\left({\bf v}_i - \overline {\bf v}\right)^{2}\rangle/2$. Here, we use only the fluctuation of the velocity, while the collective drift motion during the depinning process is removed. We express the kinetic temperature in units of $T_0 = Q^2/4\pi\epsilon_0 a$.

The kinetic temperature at the depinning of our simulated 2DDP also reflects the continuity and discontinuity of the phase transitions. For the substrate with $U_0/E_0 = 0.01 $, as the driving force increases from zero, the kinetic temperature increases gradually to its maximum value when $F_d/F_0 \approx 0.018$, then decreases gradually, as shown in Fig.~5(a). This gradual variation of the kinetic temperature probably suggests that the two transitions from the pinned to the disordered plastic flow phase, then to the final moving ordered phase, are both continuous. For $U_0/E_0 = 0.05 $ in Fig.~5(b), as the driving force increases from zero, the kinetic temperature increases gradually to its maximum value when $F_d/F_0 \approx 0.033$, then decreases abruptly, suggesting that the transition from the pinned state to the plastic flow phase is continuous, while the transition from the plastic flow to the moving ordered phase is first-order. For $U_0/E_0 = 0.10 $ in Fig.~5(c), as the driving force increases to $F_d/F_0  \approx 0.021$, the kinetic temperature jumps suddenly to a higher nonzero value, increases smoothly to its maximum when $F_d/F_0 \approx 0.049$, then decreases abruptly to a value of nearly zero. This variation of the kinetic temperature suggests that the two transitions from the pinned to the plastic flow state, then to the final moving ordered state are both first-order. Note that, for Figs.~5(a-c), the kinetic temperatures of $k_BT_x$ and $k_BT_y$, due to the motion in both $x$ and $y$ directions, is clearly synchronized. Compared with the high magnitude of the collective drift velocity in only the $x$ direction as shown in Fig.~3, the small anisotropic effects on $k_BT_x$ and $k_BT_y$ in Figs.~5(a-c) are nearly negligible.

The kinetic temperature we obtain for the substrate with $U_0/E_0 = 0.25 $ exhibits strong anisotropic effects, as shown in Fig.~5(d). Clearly, the variations in $k_BT_x$ and $k_BT_y$ are no longer synchronized, unlike what was shown in Figs.~5(a-c). When the driving force increases from zero to $F_d/F_0 \approx 0.104 $, $k_BT_x$ immediately jumps to its maximum value of $ k_BT_x / T_0 \approx 0.0055$, while $k_BT_y$ remains almost unchanged. The phase transition at this point is reasonably first-order. Then, as the driving force increases further, $k_BT_x$ decays gradually and monotonically to its minimum value.

\section{IV.~Summary}

We investigate the depinning dynamics of 2DDP under 1DPS using Langevin dynamical simulations. Note, the depinning dynamics have been investigated in many over-damped systems, however, far less is known about what happens when the mass or inertial effects play an important role, whereas dusty plasma is an ideal system to study such effects. In our study here, various diagnostics are calculated, such as the static structural measures of the sixfold coordinated particles $P_6$, the collective drift velocity $V_x$, the kinetic temperature, and the hysteresis of $V_x$ while the driving force increases and decreases monotonically. Similar to the depinning dynamics in other physical systems, we find that there are typically three different states, which are the pinned, disordered plastic flow, and moving ordered states. 

From our simulation results, we find that the depth of the substrate can change the properties of the depinning phase transitions. When the depth of the substrate is shallow, there are two continuous phase transitions. When the depth of the potential well is slightly higher, the phase transition from the pinned to the disordered plastic flow state is continuous; however, the phase transition from the disordered plastic flow state to the moving ordered state becomes discontinuous. When the substrate is even deeper, the phase transition from the pinned to the disordered plastic flow state also changes to discontinuous. When the substrate is further deepened, as the driving force increases, the pinned state jumps directly to the moving ordered state, and the disordered plastic flow state completely disappears.

\section{Acknowledgments}

Work in China was supported by the National Natural Science Foundation of China under Grant No. 11875199, the 1000 Youth Talents Plan, startup funds from Soochow University, and the Priority Academic Program Development (PAPD) of Jiangsu Higher Education Institutions. Work at LANL was supported by the US Department of Energy through the Los Alamos National Laboratory. Los Alamos National Laboratory is operated by Triad National Security, LLC, for the National Nuclear Security Administration of the U. S. Department of Energy (Contract No. 892333218NCA000001).


\begin{thebibliography}{16}


\bibitem{Reichhardt:2017}
  C.~Reichhardt and C.~J.~O.~Reichhardt,
  Depinning and nonequilibrium dynamic phase of particles assemblies driven over random and ordered substrates: A review,
  Rep. Prog. Phys. {\bf 80}, 026501~(2017).
\bibitem{Reichhardt:1998}
  C.~Reichhardt and C.~J.~Olson,and Franco Nori,
  Nonequilibrium dynamic phases and plastic flow of driven vortex lattices in superconductors
with periodic arrays of pinning sites,
  Phys. Rev. B {\bf 58}, 6534-64~(1998).
\bibitem{Harada:1996}
  K.~Harada, O.~Kamimura, H.~Kasai, T.~Matsuda, A.~Tonomura, V.~V.~Moshchalkov,
  Direct observation of vortex dynamics in superconducting films with regular arrays of defects,
  Science {\bf 274}, 1167-70~(1996).
\bibitem{Bohlein:2012}
  T.~Bohlein, J.~Mikhael, and C.~Bechinger,
  Observation of kinks and antikinks in colloidal monolayers driven across ordered surfaces,
  Nat. mater.  {\bf 11}, 126-30~(2012).
\bibitem{Williams:1991}
  F.~I.~B.~Williams, P.~A.~Wright, R.~G.~Clark, E.~Y.~Andrei, G.~Deville, D.~C.~Glattli, O.~Probst, B.~Etienne, C.~Dorin, C.~T.~Foxon, and J.~J.~Harris,
  Conduction threshold and pinning frequency of magnetically induced Wigner solid,
  Phys. Rev. Lett.  {\bf 66}, 3285-8~(1991).
\bibitem{Reichhardt:2003}
  C.~Reichhardt, C.~J.~O.~Reichhardt, I.~Martin, and A.~R.~Bishop,
  Dynamical ordering of driven stripe phases in quenched disorder,
  Phys. Rev. Lett.  {\bf 90}, 026401~(2003).
\bibitem{Sengupta:2010}
  A.~Sengupta, S.~Sengupta, and G.~I.~Menon,
  Driven disordered polymorphic solids: phases and phase transitions, dynamical coexistence and peak effect anomalies,
  Phys. Rev. B  {\bf 81}, 144521~(2010).
\bibitem{Schwarz:2001}
  J.~M.~Schwarz and Daniel S.~Fisher,
  Depinning with dynamic stress overshoots: Mean Field Theory,
  Phys. Rev. Lett. {\bf 87}, 096107~(2001).


\bibitem{Thomas:1996}
  H.~M.~Thomas and G.~E.~Morfill,
  Melting dynamics of a plasma crystal,
  Nature (London) {\bf 379}, 806~(1996).
\bibitem{L:1996}
  L.~I, W.~T.~Juan, C.~H.~Chiang, and J.~H.~Chu,
  Microscopic particle motions in strongly coupled dusty plasmas,
  Science {\bf 272}, 5268~(1996).
\bibitem{Konopka:2000}
  U.~Konopka, G.~E.~Morfill, and L.~Ratke,
  Measurement of the interaction potential of microspheres in the sheath of a rf discharge,
  Phys. Rev. Lett. {\bf 84}, 891~(2000).
\bibitem{Merlino:2004}
  R.~Merlino and J.~Goree,
  Dusty plasmas in the laboratory, industry, and space,
  Phys. Today {\bf 57}(7), 32~(2004).
\bibitem{Fortov:2005}
  V.~E.~Fortov, A.~V.~Ivlev, S.~A.~Khrapak, A.~G.~Khrapak, and G.~E.~Morfill,
  Complex (dusty) plasmas: current status, open issues, perspectives,
  Phys. Rep. {\bf 421}, 1~(2005).
\bibitem{Morfill:2009}
  G.~E.~Morfill and A.~V.~Ivlev,
  Complex plasmas: an interdisciplinary research field,
  Rev. Mods. Phys. {\bf 81}, 1353~(2009).
\bibitem{Bonitz:2010}
  M.~Bonitz, C.~Henning, and D.~Block,
  Complex plasmas: a laboratory for strong correlations,
  Rep. Prog. Phys. {\bf 73}, 066501~(2010).
\bibitem{Melzer:2013}
  A.~Melzer, A.~Schella, J.~Schablinski, D.~Block, and A.~Piel,
  Analyzing the liquid state of two-dimensional dust clusters: The instantaneous normal mode approach,
  Phys. Rev. E {\bf 87}, 033107~(2013).
\bibitem{Thomas:2016}
  E.~Thomas, Jr., U.~Konopka, R.~Merlino, and M.~Rosenberg,
  Initial measurements of two- and three-dimensional ordering, waves, and plasma filamentation in the Magnetized Dusty Plasma Experiment,
  Phys. Plasmas {\bf 23}, 055701~(2016).


\bibitem{Li:2018}
  W.~Li, D.~Huang, K.~Wang, C.~Reichhardt, C.~J.~O.~Reichhardt, M.~S.~Murillo, and Y.~Feng,
  Phonon spectra of two-dimensional liquid dusty plasmas on a one-dimensional periodic substrate,
  Phys. Rev. E {\bf 98}, 063203~(2018).
\bibitem{Wang:2018}
  K.~Wang, W.~Li, D.~Huang, C.~Reichhardt, C.~J.~O.~Reichhardt, M.~S.~Murillo, and Y.~Feng,
  Structures and diffusion of two-dimensional dusty plasmas on one-dimensional periodic substrates,
  Phys. Rev. E {\bf 98}, 063204~(2018).
\bibitem{Li:2020}
  W.~Li, C.~Reichhardt, C.~J.~O.~Reichhardt, M.~S.~Murillo, and Y.~Feng,
  Oscillation-like diffusion of two-dimensional liquid dusty plasmas on one-dimensional periodic substrates with varied widths,
  Phys. Plasmas {\bf 27}, 033702~(2020).
\bibitem{Feng:2011}
  Y.~Feng, J.~Goree, B.~Liu, and E.~G.~D.~Cohen,
  Green-Kubo relation for viscosity tested using experimental data for a two-dimensional dusty plasma,
  Phys. Rev. E  {\bf 84}, 046412~(2011)
\bibitem{Qiao:2014}
  K.~Qiao, J.~Kong, J.~Carmona-Reyes, L.~S. Matthews,and T.~W.~Hyde,
  Mode coupling and resonance instabilities in quasi-two-dimensional dust clusters in complex plasmas,
  Phys. Rev. E  {\bf 90}, 033109~(2014)
\bibitem{Feng:2008}
  Y.~Feng, J.~Goree, and B.~Liu,
  Solid superheating observed in two-dimensional strongly coupled dusty plasma,
  Phys. Rev. Lett.  {\bf 100}, 205007~(2008).
\bibitem{Hartmann:2014}
  P.~Hartmann, A.~Z.~Kov\'acs, A.~M.~Douglass, J.~C.~Reyes, L.~S.~Matthews, and T.~W.~Hyde,
  Slow plastic creep of 2D dusty plasma solids,
  Phys. Rev. Lett.  {\bf 113}, 025002~(2014).
\bibitem{Melzer:1996}
  A.~Melzer, A.~Homann, and A.~Piel,
  Experimental investigation of the melting transition of the plasma crystal,
  Phys. Rev. E  {\bf 53}, 2757~(1996).
\bibitem{Chan:2007}
  C.-L.~Chan and L.~I,
  Microstructural evolution and irreversibility in the viscoelastic response of mesoscopic dusty-plasma liquids,
  Phys. Rev. Lett.  {\bf 98}, 105002~(2007)
\bibitem{Feng:2010}
  Y.~Feng, J.~Goree, and B.~Liu,
  Viscoelasticity of 2D liquids quantified in a dusty plasma experiment,
  Phys. Rev. Lett.  {\bf 105}, 025002~(2010)
\bibitem{Thomas:2004}
  E.~Thomas, Jr., J.~D.~Williams, and J.~Sliver,
  Application of stereoscopic particle image velocimetry to studies of transport in a dusty (complex) plasma,
  Phys. Plasmas {\bf 11}, L37~(2004).
\bibitem{LiF:2009}
  Y.~F.~Li, U.~Konopka, K.~Jiang, T.~Shimizu, H.~Höfner, H.~M.~Thomas,
  Removing dust particles from a large area discharge,
  Appl. Phys. Lett. {\bf 94}, 081502~(2009).
\bibitem{LiF:2010}
  Y.~F.~Li, W.~G.~Zhang, J.~X.~Ma, K.~Jiang, H.~M.~Thomas, and G.~E.~Morfill,
  Traveling electric field probed by a fine particle above voltage-modulated strips in a striped electrode device,
  Phys. Plasmas {\bf 17}, 033705~(2010).
  
\bibitem{Melzer:2000}
A.~Melzer, S.~Nunomura, D.~Samsonov, Z.~W.~Ma, and J.~Goree, Laser-excited Mach cones in a dusty plasma crystal, Phys. Rev. E {\bf 62}, 4162~(2000).
  
\bibitem{Li:2019}
  W.~Li, K.~Wang, C.~Reichhardt, C.~J.~O.~Reichhardt, M.~S.~Murillo, and Y.~Feng,
  Depinning dynamics of two-dimensional dusty plasmas on a one-dimensional periodic substrate,
  Phys. Rev. E {\bf 100}, 033207~(2019).

\bibitem{Reichhardt:1997}
  C.~Reichhardt, C.~J.~Olson, and Franco Nori,
  Dynamic phases of vortices in superconductors with periodic pinning,
  Phys. Rev. Lett. {\bf 78}, 2648-51~(1997).
\bibitem{Reichhardt:2005}
  C.~Reichhardt and C.~J.~O.~Reichhardt,
  Pinning and dynamics of colloids on one-dimensional periodic potentials,
  Phys. Rev. E {\bf 72}, 032401~(2005).


\bibitem{Ohta:2000}
  H.~Ohta and S.~Hamaguchi,
  Molecular dynamics evaluation of self-diffusion in Yukawa systems,
  Phys. Plasmas {\bf 7}, 4506~(2000).
\bibitem{Sanbonmatsu:2001}
  K.~Y.~Sanbonmatsu and M.~S.~Murillo,
  Shear viscosity of strongly coupled Yukawa systems on finite length scales,
  Phys. Rev. Lett. {\bf 86}, 1215~(2001).
\bibitem{Kalman:2004}
  G.~J.~Kalman, P.~Hartmann, Z.~Donk\'o, and M.~Rosenberg,
  Two-dimensional Yukawa liquids: correlation and dynamics,
  Phys. Rev. Lett. {\bf 92}, 065001~(2004).

\bibitem{Donko:2010}
  Z.~Donk\'o, J.~Goree, and P.~Hartmann,
  Viscoelastic response of Yukawa liquids,
  Phys. Rev. E {\bf 81}, 056404~(2010).


\bibitem{Liu:2003}
  B.~Liu, K.~Avinash, and J.~Goree,
  Transverse optical mode in a one-dimensional Yukawa chain,
  Phys. Rev. Lett. {\bf 91}, 255003~(2003).
\bibitem{FengY:2008}
  Y.~Feng, B.~Liu, and J.~Goree,
  Rapid heating and cooling in two-dimensional Yukawa systems,
  Phys. Rev. E {\bf 78}, 026415~(2008).
\bibitem{van:1982}
  W.~F.~van Gunsteren and H.~J.~C.~Berendsen,
  Algorithms for Brownian dynamics,
  Mol. Phys. {\bf 45}, 637~(1982).
\bibitem{Hartmann:2005}
  P.~Hartmann, G.~J.~Kalman, Z.~Donk\'o, and K.~Kutasi,
  Equilibrium properties and phase diagram of two-dimensional Yukawa systems,
  Phys. Rev. E {\bf 72}, 026409~(2005).


\bibitem{Thorel:1973}
  P.~Thorel, R.~Kahn, Y.~Simon, and D.~Cribier,
  Fabrication et\'etude d'un monocristal de vortex dans le niobium supracconducteur,
  J. Phys. {\bf 34}, 447~(1973).
\bibitem{Shi:1991}
  A.~-C.~Shi and A.~J.~Berlinsky,
  Pinning and I-V characteristics of a two-dimensional defective flux-line lattice,
  Phys. Rev. Lett. {\bf 67}, 1926~(1991).
\bibitem{Koshelev:1994}
  A.~E.~Koshelev and V.~M.~Vinokur,
  Dynamic melting of the vortex lattice,
  Phys. Rev. Lett. {\bf 73}, 3580~(1994).
\bibitem{Reichhardt:2015}
  C.~Reichhardt, D.~Ray, and C.~J.~O.~Reichhardt,
  Collective transport properties of driven skyrmions with random disorder,
  Phys. Rev. Lett. {\bf 114}, 217202~(2015).

\bibitem{SM}
See the Supplemental Materials for the typical trajectories from our simulated 2DDP under 1DPS.

\bibitem{Ooyen:1992}
  A.~van Ooyen and J.~van Pelt,
  Phase transitions, hysteresis and overshoot in developing neural networks,
  In: I.~Aleksander and J.~Taylor (Eds.), Artificial Neural Networks, 2, Proc. ICANN'92, Elsevier, Amsterdam, pp. 907-910.~(1992).
\bibitem{Schwarz:2003}
  J.~M.~Schwarz and Daniel S.~Fisher,
  Depinning with dynamic stress overshoots: A hybrid of critical and pseudohysteretic behavior,
  Phys. Rev. E {\bf 67}, 021603~(2003).



\end{thebibliography}

\end{document}